\documentclass[12pt,preprint]{aastex}

\begin{document}

\title{Physical Properties of the AXP 4U 0142+61 from X-ray Spectral Analysis}

\author{Tolga G\"uver\altaffilmark{1}, Feryal \"Ozel\altaffilmark{2}
  and Ersin G\"o\u{g}\"u\c{s}\altaffilmark{3}}

\altaffiltext{1}{Istanbul University, Science Faculty, Astronomy \&
  Space Sciences Department, Beyaz\i t, Istanbul, 34119}
\altaffiltext{2}{University of Arizona, Department of Physics, 1118
  E. 4th St., Tucson, AZ 85721} 
\altaffiltext{3}{Sabanc\i~University,
  Faculty of Engineering Natural Sciences, 34956 Turkey}

\begin{abstract}

We analyze archival {\it Chandra} and XMM-{\it Newton} data of
4U~0142$+$61 within the context of the Surface Thermal Emission and
Magnetospheric Scattering model. We show that 4U~0142$+$61 spectrum
can be fit very well with this physical model that contains only four
parameters. The system parameters can be tightly constrained from the
fits, yielding a surface magnetic field strength of $B=(4.75 \pm 0.02)
\times 10^{14}$~G, a surface temperature of $kT = 0.309 \pm 0.001$~keV
and a scattering optical depth of a few in the magnetosphere.  These
values do not vary between observations due to the stability of the
source within the window of the observations. The detailed fits yield
$\chi^2$ values that are statistically much better than the
traditionally employed blackbody+power-law and two blackbody fits. The
spectroscopically measured surface magnetic field strength is higher
than, but within, the theoretical uncertainties of the value inferred
from the dipole spindown formula.

\end{abstract}


\section{Introduction}

Anomalous X-ray Pulsars (AXPs) and Soft Gamma Repeaters (SGRs) are
thought to be the observational manifestations of a class of
ultramagnetic ($B \gtrsim 10^{14}$~G) neutron stars, also called
magnetars (see Woods \& Thompson 2006 and Kaspi 2006 for a detailed
review on AXPs and SGRs). The strong magnetic fields are believed to
power the X-ray emission of these neutron stars and give rise to high
spin-down rates ($ \dot{P} \sim 10^{-11}$~s~s$^{-1}$) (Thompson \&
Duncan 1996).  Furthermore, the large reservoir of magnetic energy
associated with such fields leads to intense, super-Eddington (L
$\gtrsim$ L$_{\rm Edd}$), random bursts of X-rays or soft
gamma-rays. Indeed, observations of such powerful bursts that
typically last a fraction of a second and have been detected from all
four known SGRs and at least five out of the eight known AXPs lend
strong, albeit indirect, support for their identification as magnetars
(Gavriil \& Kaspi 2002; Kaspi et al. 2003; Woods et al. 2005).

AXPs and SGRs are all observed as point X-ray sources with
luminosities of $10^{33-36}$~erg~s$^{-1}$. Their X-ray spectra, in the
$0.5-10.0$~keV photon energy range, have so far been described by
empirical functions such as a blackbody (kT $\sim$ 0.3$-$0.6 keV) plus
a power law (with photon index $\Gamma \sim$2.5$-$4) and, less
frequently, by a sum of two blackbody functions (see, e.g., Gotthelf
\& Halpern 2005; Kaspi 2006). Within the magnetar model, the blackbody
component is attributed to the emission from the neutron star surface
that is heated by the decay of the strong magnetic field (Thompson \&
Duncan 1996). The power law component, on the other hand, is thought
to be magnetospheric in origin and is widely used to obtain a better
representation of the X-ray spectra.

We have recently developed a physical model of emission from a
magnetar that takes into account processes in its atmosphere as well
as in its magnetosphere. The Surface Thermal Emission and
Magnetospheric Scattering (STEMS) model is based on the radiative
equilibrium atmosphere calculations presented in \"Ozel (2003) but
also includes the effects of magnetospheric scattering of the surface
radiation as discussed in Thompson Lyutikov and Kulkarni (2002),
Lyutikov \& Gavriil (2006) and G\"uver, \"Ozel \& Lyutikov (2007). Our
models predict strong deviations from a Planckian spectrum, with a
hard excess that depends on the surface temperature as well as the
magnetic field strength (\"Ozel \& G\"uver 2007), and weak absorption
lines due to the proton cyclotron resonance. Both the atmospheric
processes and the magnetospheric scattering play a role in forming
these spectral features and especially in reducing the equivalent
widths of the cyclotron lines.

With the first successful application of this model (G\"uver et al.\
2007), we fit the spectrum of the AXP XTE J1810$-$197, a transient
source whose flux showed more than two orders of magnitude variation
during the three years it has been monitored (Gotthelf \& Halpern
2006). In contrast, 4U~0142$+$61 is the brightest and historically a
stable AXP. Following its detection with {\it Uhuru}, an EXOSAT
campaign revealed its neutron star nature by the discovery of its 8.7
s periodicity (Israel et al. 1994).  Multiple X-ray observations of
the source showed a long epoch of nearly constant flux levels as well
as a relatively hard X-ray spectrum (Juett et al. 2002; Patel et
al. 2003; G\"ohler, Wilms \& Staubert 2005). Recently, the source
exhibited SGR like bursts (Kaspi, Dib \& Gavriil 2006; Dib et
al. 2006; Gavriil et al. 2007) for the first time.

4U 0142$+$61 has also been detected in hard X-rays with INTEGRAL
(Kuiper et al. 2006, den Hartog et al. 2007). The hard X-ray spectral
component in the $20-230$~keV energy range is well described by a
power law model of index $0.79$ and the corresponding flux is $1.7
\times 10^{-10}$~erg~cm$^{-2}$s$^{-1}$ (den Hartog et al. 2007),
which exceeds by a factor of $\sim$2 the unabsorbed 2-10 keV flux. It
is noteworthy that the extrapolation of the power-law component
towards lower photon energies yields flux levels that contribute
significantly to the soft X-ray flux at 7-10 keV range. Furthermore,
the fact that the hard X-ray component is pulsed and is in phase with
soft X-rays (Kuiper et al. 2006) points to a connection between the
hard and soft components. Rea et al. (2007) attempted to model the
combined soft and hard X-ray spectrum with a variety of empirical
functions and a model that treats resonant scattering in the
magnetosphere and showed that some of these empirical models were
feasible.

Durant \& van Kerkwijk (2006a) measured the galactic column density to
some of the AXPs, using the individual absorption edges of the
elements O, Fe, Ne, Mg, and Si. They found the column density to
4U~0142$+$61 to be $(0.64 \pm 0.07) \times 10^{22}\; \rm{cm}^{-2}$, a factor of
1.4 lower than the value inferred from the blackbody plus power-law
fits. Using the red clump stars (core-helium burning giants) in the
direction of the source to measure the variation of the reddening with
distance and extinction, Durant \& van Kerkwijk (2006b) also
determined the distance of the source as 3.6~kpc.

In this paper, we analyze archival {\it Chandra} and XMM-{\it Newton}
data of 4U~0142$+$61 within the context of the STEMS model and obtain
physical system parameters by performing detailed fits to the soft
X-ray spectra of this source. In the next section, we describe our
physical model. In Section 3, we present the data and the fit
results. We conclude in Section 4 with a discussion of our results and
their implications.

\section{The Surface Thermal Emission and Magnetospheric Scattering
  Model}

The spectrum of a magnetar is molded by its atmosphere and its
magnetosphere. In the ionized, highly magnetic neutron star
atmospheres, polarization-mode dependent transport of radiation that
includes absorption, emission, and scattering processes determine the
continuum spectrum (see e.g. \"Ozel 2001, 2003; Lai \& Ho
2003). Furthermore, the interaction of the photons with the protons in
the plasma gives rise to an absorption feature at the proton cyclotron
energy

\begin{equation}
E_p = 6.3 \; \left(\frac{B}{10^{15}\;{\rm G}}\right) \; {\rm keV}.
\end{equation}

This absorption feature is weakened by the vacuum polarization
resonance, which also leads to an enhanced conversion between photons
of different polarization modes as they propagate through the
atmosphere.

In the magnetospheres, currents supporting the ultrastrong magnetic
fields can lead to enhanced charge densities (Thompson, Lyutikov, \&
Kulkarni 2002), which can reprocess the surface radiation through
resonant cyclotron scattering (Lyutikov \& Gavriil 2006; G\"uver,
\"Ozel, \& Lyutikov 2007).  We calculate this effect using the Green's
function approach described in Lyutikov \& Gavriil (2006) assuming
that the magnetosphere is spherically symmetric and the field
strength follows a 1/$\rm{r}^{3}$ dependence.

We have developed a spectral model that includes these relevant
mechanisms that take place on the magnetar surface and its
magnetosphere and depends only on four physical parameters. The first
two parameters, the surface magnetic field strength $B$ and
temperature $T$, describe the conditions found on the neutron star
surface.  The third parameter denotes the average energy of the
charges $\beta = v_e/c$ in the magnetosphere, while the last parameter
is related to the density $N_e$ of such charges and indicates the
optical depth to resonant scattering by

\begin{equation}
\tau = \sigma \int N_e dz. 
\end{equation}

Here, $\sigma$ is the cross-section for resonant cyclotron scattering.
We also assume a fixed value for the gravitational acceleration on the
neutron star surface of $1.9\times10^{14}$~cm~s$^{-2}$, obtained for
reasonable values of the neutron star mass and radius.

We calculated model X-ray spectra (in the 0.05 - 9.8 keV range) by
varying model parameters in suitable ranges that are in line with the
physical processes we incorporated into the models: surface
temperature $T=0.1$ to $0.6$~keV, magnetic field $B=5\times10^{13}$ to
$3\times10^{15}$ G, electron velocity $\beta = 0.1$ to 0.5, and
optical depth in the magnetosphere $\tau =1$ to 10.  From the set of
calculated spectra, we created a table model which we use within the
X-ray spectral analysis package XSPEC (Arnaud 1996) to model the X-ray
spectra of 4U~0142+61.

\section{Observations \& Data Analysis}

In Table~\ref{obs}, we present the list of the archival pointed X-ray
observations of 4U~0142$+$61 that we analysed in this study. {\it
Chandra} observations were calibrated using
CIAO\footnote{http://cxc.harvard.edu} v.3.4 and CALDB 3.3.0.1.  For
the XMM-Newton observations we used the Science Analysis Software
(SAS) v.7.0.0 and the latest available calibration files.  The
XMM-Newton observation in 2002 was excluded from earlier studies
(G\"ohler, Wilms \& Staubert 2005) because it was partially affected
by the high energy particle background.  We were able to eliminate the
segments with a high background, and were able to utilize an effective
exposure of 1.9 ks out of the $3.4$ ks. We used only EPIC-PN data of
each XMM-Newton observation.

For the small window mode XMM-Newton observations, we extracted source
spectra from a circle centered on the source with a radius of
32\arcsec and the background from a source free region with a radius
of 50\arcsec. We extracted the source region from the CC mode Chandra
observation using a rectangular region centered on the source with
sizes 8 $\times$ 2 \arcsec, and used as the background region from
this dataset a source-free region with similar sizes. For the
XMM-Newton observations in the fast-timing mode, we extracted the
source spectrum from a rectangular region of 9.5 pixels centered on
the source, and used a background spectrum with similar sizes from a
source-free region on the CCD. To create the response and anciallary
response files, we used mkacisrmf, mkarf and epproc tasks for Chandra
and XMM-Newton datasets, respectively. We rebinned XMM-{\it Newton}
spectra such that each energy bin contains at least 50 counts without
oversampling the energy resolution of the instrument. To account for
the calibration uncertainties we have also included a 2\% systematic
error in all fits.

The spectral analysis was performed using the XSPEC 11.3.2.t (Arnaud
1996). We assumed a fiducial gravitational redshift correction of 0.2,
which corresponds to a neutron star with mass 1.4 $M_{\sun}$ and $R =
13.8$~km. We calculate the fluxes for the $0.5-8.0$~keV energy range
and quote errors for 90\% confidence level. For the calculation of
galactic column density, we have used Anders \& Grevesse (1989) solar
abundances.

\subsection{Results of Spectral Modeling}

In our analysis, we take into account the contribution of the hard
X-ray emission to the 0.5 - 8.0 keV spectra by adding a power-law
component with frozen parameters given by den Hartog et al. (2007). In
doing so, we assume that this hard power-law component extends down to
the soft X-ray band without a break and thus has a non-negligible
contribution to the overall flux above 6.5 keV. In addition, we take
into account the independent results of Durant \& van Kerkwijk (2006a)
to evaluate the performance of the models at low energies.

The spectral properties of 4U~0142$+$61 do not vary significantly
throughout the four years spanned by the observations.  We, therefore,
first fit all XMM-Newton EPIC-PN spectra simultaneously in order to
better constrain model parameters. Note that we excluded Chandra
ACIS-S observation from the simultaneous fit to avoid any systematic
uncertainties due to different calibration schemes. We obtained an
excellent fit to data, $\chi^{2}_{\nu}$ = 0.949 for 1534 degrees of
freedom (d.o.f.), with flat residuals. The data, best-fit model, and
the residuals are shown in Figure~1.

The fit provides tight constrains on the model parameters: the surface
temperature, $kT = 0.309 \pm 0.001$~keV, the surface magnetic field
strength $B=(4.75 \pm 0.02) \times 10^{14}$~G, the optical depth to
scattering in the magnetosphere, $\tau = 3.57 \pm 0.03$, and a thermal
particle velocity in the magnetosphere $\beta = 0.417 \pm 0.002$.

For the hydrogen column density, we obtain $N_{\rm{H}} = (0.566 \pm
0.002) \times 10^{22} \rm{cm}^{-2}$, which is in good agreement with
the value (of $N_{\rm{H}}=0.64 \pm0.07 \times 10^{22}\ \rm{cm}^{-2}$)
found by Durant \& van Kerkwijk (2006a).  If we, instead, demand an
exact correspondence with the latter value by freezing the column
density at $0.64 \times 10^{22}\; \rm{cm}^{-2}$, we obtain a somewhat
poorer fit ($\chi^{2}_{\nu}$ = 1.361 for 1535 d.o.f.).

We also fit each spectrum individually with the STEMS model.  We find
that our model produces excellent fits to all individual spectra.  In
Table \ref{restfl}, we present the results of the individual spectral
fits. The obtained values of the parameters are consistent with each
other within $1-\sigma$, as well as with the results of the
simultaneous fit.

For comparison, we have also used the empirical blackbody plus
power-law model to fit all XMM-{\it Newton} data simultaneously (still
allowing for a contribution from the extension of the hard X-ray
power-law component). The result is acceptable within the context of
X-ray spectroscopy ($\chi^{2}_{\nu}$ of 1.349 for 1532 d.o.f.);
however, the residuals, at especially $\lesssim 2$ keV, are not flat
(see upper panel of Figure \ref{all_bbpl}) and do not capture the
characteristics of the spectrum. Correspondingly, the $\chi^2_{\nu}$
value is worse than that we obtain for our STEMS fits even though the
STEMS model has two fewer parameters than the blackbody plus power-law
model.\footnote{Note that to describe each data set, the blackbody
  plus power-law model requires one fewer parameter than the STEMS
  model. However, we allow the normalizations of the blackbody and the
  power-law components to vary independently for the description of
  each data set, resulting in a total of 11 free parameters in the
  simultaneous fit to four data sets. On the other hand, the STEMS
  model has 1 normalization per data set, which yields a total number
  of 9 free parameters for the 4 data sets.} For the blackbody plus
power-law fit, we obtain model parameters of $N_{\rm H} = 1.001 \pm
0.002 \times 10^{22} \; \rm{cm}^{-2}$, a blackbody temperature of
0.431$\pm0.001$ keV and a photon index of 3.94$\pm0.01$ (see Figure
\ref{all_bbpl} upper panel). Note that the column density is 1.6 times
(92$\sigma$) higher than the value reported by (Durant \& van Kerkwijk
2006a) through a different and spectral model-independent analysis.
Because it shows this large disagreement, we also tried a fit where
the column density is fixed at the latter value. The resulting fit is
unacceptable, with $\chi^2_{\nu}$ = 3.83 for 1533 d.o.f., ${\rm kT} =
0.40\pm0.01$~keV, and photon index $\Gamma = 2.83\pm0.01$. We show
this fit in the lower panel of Figure \ref{all_bbpl}. The large
discrepancy between the column density value of the blackbody plus
power-law analysis and that of Durant \& van Kerkwijk (2006a) is
likely to be due to the fact that the power-law component is
artificial and, because of steep photon index values, needs to be
attenuated significantly at low energies, requiring large $N_{\rm H}$
values.

We have also attempted to fit the combined X-ray spectra of 4U~0142+61
with two blackbodies and the hard X-ray power-law.  The resulting
statistics ($\chi^{2}_{\nu}$ = 1.19 for 1532 d.o.f.) are acceptable,
slightly better than blackbody plus power-law fits, but perform poorly
compared to the physical STEMS model ($\Delta\chi^2$ = 354.8,
$\Delta\nu$ = 2).  These fits do better than the blackbody plus
power-law fits in obtaining $N_{\rm{H}}$ values (0.559$\pm0.004
\times10^{22} \rm{cm}^{-2}$) closer to those determined by Durant \&
van Kerkwijk (2006a). However, there are positive residuals that rise
systematically above 5.0~keV, indicating that the two blackbody fits
do not capture the observed hardness of the 4U~0142$+$61 spectra
fully, even with the contribution from the hard X-ray power-law
component.  For these fits, we obtain parameter values at $kT_{1} =
0.349 \pm0.002$ and $kT_{2} = 0.719 \pm 0.003$~keV, consistent with
those found by den Hartog et al. (2007).

\section{Discussion}

\begin{figure*}
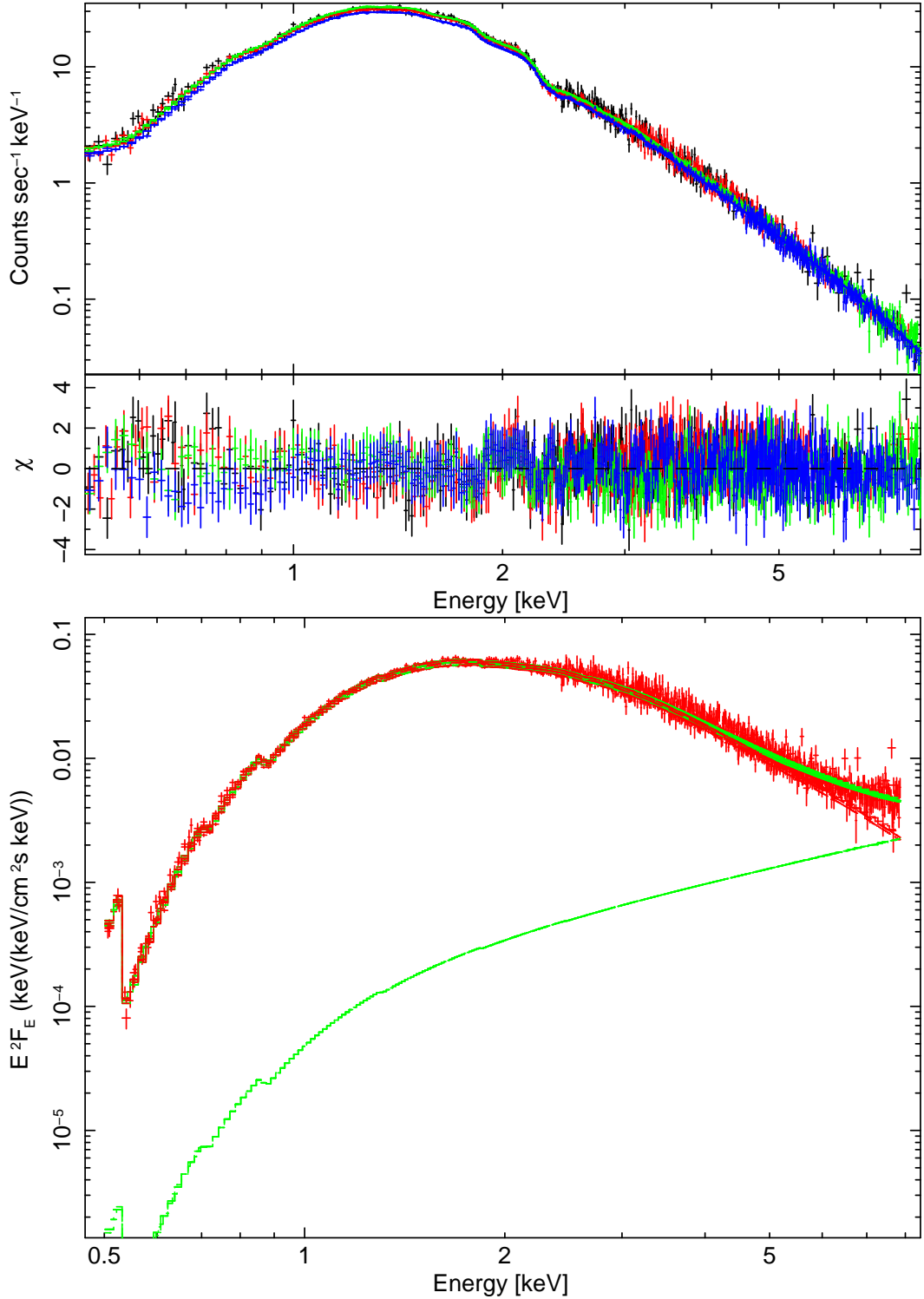

\centering
   \includegraphics[angle=270,scale=0.60]{f1a.ps}
   \includegraphics[angle=270,scale=0.60]{f1b.ps}
   \caption{Simultaneous fit of the Surface Thermal Emission and
     Magnetospheric Scattering model to the four data sets of the
     X-ray spectra of 4U~0142$+$61 given in Table~1. In the lower
     panel, the E$^{2}$F$_{E}$ spectra shows the effects of the extrapolated
     hard X-ray component on the soft X-ray spectra.}
\label{all}
\end{figure*}

\begin{figure*}
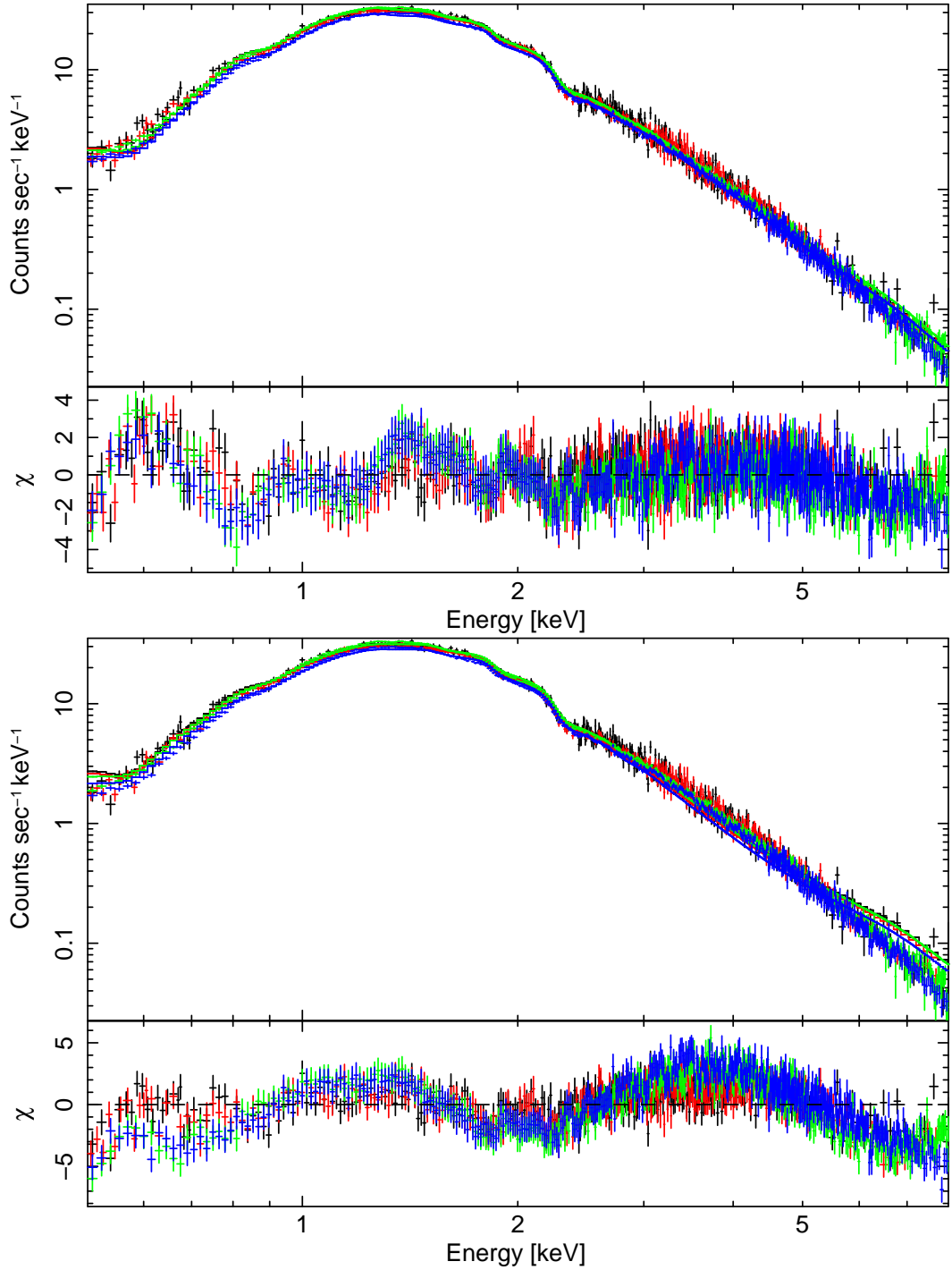

\centering 
\includegraphics[angle=270,scale=0.60]{f2a.ps}
\includegraphics[angle=270,scale=0.60]{f2b.ps}
   \caption{Simultaneous blackbody plus power-law model fit to the
   X-ray spectra of 4U~0142$+$61. The upper panel shows the fit when
   the hydrogen column density $N_{\rm H}$ is allowed to vary, while
   the lower panel corresponds to $N_{\rm H}$ frozen at $0.64 \times
   10^{22} \; \rm{cm}^{-2}$, the value measured independently by
   Durant \& van Kerkwijk (2006a). }
\label{all_bbpl}
\end{figure*}

In this paper, we analyzed the archival XMM and {\it Chandra} data on
4U~0142$+$61 and showed that the X-ray spectrum of this magnetar can
be fit very well with the Surface Thermal Emission and Magnetospheric
Scattering model. The model contains only four physical parameters,
which can be tightly constrained from the spectral fits. The residuals
are flat over the energy range of the observations, as shown in Figure
\ref{all}, indicating the ability of our model to capture the shape of
the continuum in the whole energy range.

Our model also allows us to determine the physical properties of the
neutron star. We show in Figure~3 the confidence contours we obtain
for the surface magnetic field strength, temperature, the scattering
optical , the particle velocity in the magnetosphere and galactic
column density. We obtain tight constraints for these parameters. In
particular, the narrow contours for the magnetic field strength and
the temperature are because of the fact that they both cause small but
detectable variations in the X-ray spectra, which, combined with high
quality spectra and a large number of data points, pin down the values
of these parameters. The scattering optical depth contributes to both
the hardness of the model spectra and to attenuating proton cyclotron
features and is also well-constrained by the observations.

The measured values of temperature, surface magnetic field strength,
and magnetospheric parameters remain constant within statistical
uncertainty for each data set because of the stability of the source
over the observed period. Figure~3 also shows that for most
parameters, the errors are uncorrelated.

Comparison of the STEMS model to data also gives us the chance to
probe the magnetospheres of magnetars. The scattering optical depth of
3.57 we obtain in our analysis corresponds to a charged particle
density in the magnetosphere that is approximately 3$\times10^{5}$
times higher than the Goldreich-Julian density for 4U 0142$+$61,
(using the inferred dipole magnetic field strength and the spin period
of the source). Note that the magnetospheric parameteres we report
here differ from those given in the resonant cyclotron scattering model
of Rea et al.\ (2007) both because that analysis fitted a different
power-law index for the hard X-ray component and because it does not
take into account the atmospheric effects but uses a canonical
blackbody.

We have determined the surface magnetic field strength of 4U~0142$+$61
as $4.75\times10^{14}$~G. This value is quite close to the dipole
field strength $1.3\times10^{14}$~G (Gavriil \& Kaspi 2002), obtained
from the spindown rate of this source using the dipole spindown
formula but is not in exact agreement as in the case of XTE
J1810$-$197, where the two field strengths are equal (G\"uver et al.\
2007). The small difference is likely due to the fact that our
spectroscopic measurements are sensitive to the field strength at the
surface of the neutron star, while the dipole spin-down method
measures the magnetic field strength at the light cylinder. In
addition, the dipole spindown formula assumes a fiducial angle between
the rotation and the magnetic axes that is not expected to be accurate
on a source-by-source basis.  Finally, recent (see, e.g., Spitkovsky
2006) numerical calculations on the structure of a rotating neutron
star magnetosphere show violations of the vacuum
assumption. Nevertheless, obtaining a surface field strength that is
close to the dipole strength is a further indication of the
reliability of the measurements and the magnetar strength fields
present in 4U~0142$+$61.

Finally, we calculate the area of the emitting region, assuming a
gravitational redshift of 0.2, a distance of 3.6 kpc (Durant \& van
Kerkwijk 2006b), and using the flux and the spectroscopically
determined surface temperature. We obtain a radius of 10.8~km that
does not vary between observations. Such a radius indicates that the
emission arises from roughly three-quarters of the whole neutron star
surface. This large surface area is suggestive: interestingly, the
X-ray pulsed fraction (Woods \& Thompson 2006) of 4U 0142+61 is 3.6\%,
which is the least among all the AXPs and SGRs.

\acknowledgements 

We thank Dr. Keith Arnaud for his help during the creation of the
XSPEC table model. This work makes use of observations obtained with
XMM-Newton, an ESA science mission with instruments and contributions
directly funded by ESA Member States and NASA. F.\ \"O. acknowledges
support from NSF grant AST-0708640 and from a Turkish Science and
Technology Council Visiting Faculty fellowship. E.G. acknowledges
partial support from the Turkish Academy of Sciences through grant
E.G/T\"UBA-GEB\.IP/2004-11.  T.G. and E.G. acknowledge EU FP6 Transfer
of Knowledge Project "Astrophysics of Neutron Stars"
(MTKD-CT-2006-042722).

\clearpage

\begin{deluxetable}{cccccc}
\tablecolumns{6}
\tablewidth{475pt}
\tablecaption{Observations used for this study.}
\tablehead{Satellite  & Detector & Mode	   & Exp. Time (ks) & Obs ID  	   & Obs Date	}
\startdata
Chandra		      & ACIS-S   & CC		 & 5.94	    & 724	   & May 21 2000   \\
XMM-Newton  	      & EPIC-PN  & Small Window  & 1.9      & 0112780301   & Feb 13 2002   \\
XMM-Newton  	      & EPIC-PN  & Small Window  & 4.0	    & 0112781101   & Jan 24 2003   \\
XMM-Newton  	      & EPIC-PN  & Fast Timing   & 35.78    & 0206670101   & Mar 01 2004   \\
XMM-Newton  	      & EPIC-PN  & Fast Timing   & 21.1     & 0206670201   & Jul 25 2004   \\
\enddata
\label{obs}
\end{deluxetable}

\begin{deluxetable}{ccccccccc}
  \tablecolumns{8} \tablewidth{520pt} \tablecaption{Results of the
    Surface Thermal Emission and Magnetospheric Scattering model. All
    the errors are given with 90\% confidence.}  \tablehead{
    Obs. Date    & $N_{\bf H}$ & Mag. Field 	 & Temp. & $\beta$ & $\tau$   & Flux\tablenotemark{a} & $\chi^{2}_{\nu}$(d.o.f.) \\
    &$(10^{22} \; \rm{cm}^{-2})$ & ($10^{14}$~G) & (keV) & (c) & & & }
  \startdata
  05/21/2000  & 0.57$\pm$0.02 & 3.96$\pm$0.31 & 0.307$\pm$0.006  & 0.46$\pm$0.02  & 5.44$\pm$0.52  & 1.92$\pm$0.20 & 1.174 (283)	\\
  02/13/2002  & 0.54$\pm$0.02 & 4.66$\pm$0.56 & 0.31$\pm$0.01    & 0.42$\pm$0.05  & 3.68$\pm$0.59  & 2.07$\pm$0.54 & 1.027 (310)	\\
  01/24/2003  & 0.55$\pm$0.02 & 5.16$\pm$0.42 & 0.31$\pm$0.01    & 0.45$\pm$0.03  & 3.16$\pm$0.33  & 2.10$\pm$0.23 & 0.999 (345)	\\
  03/01/2004  & 0.57$\pm$0.01 & 4.60$\pm$0.07 & 0.310$\pm$0.002  & 0.40$\pm$0.01  & 3.55$\pm$0.14  & 2.04$\pm$0.06 & 0.974 (402)	\\
  07/25/2004  & 0.58$\pm$0.01 & 4.67$\pm$0.16 & 0.305$\pm$0.002  & 0.43$\pm$0.01  & 3.54$\pm$0.14  & 1.98$\pm$0.07 & 0.931 (462)	\\
\enddata
\tablenotetext{a}{Unabsorbed 0.5-8.0 keV flux in units of $10^{-10}$ erg $\rm{cm}^{-2}\;\rm{s}^{-1}$}
\label{restfl}
\end{deluxetable}

\clearpage

\begin{figure*}
\centering
   \includegraphics[scale=0.40]{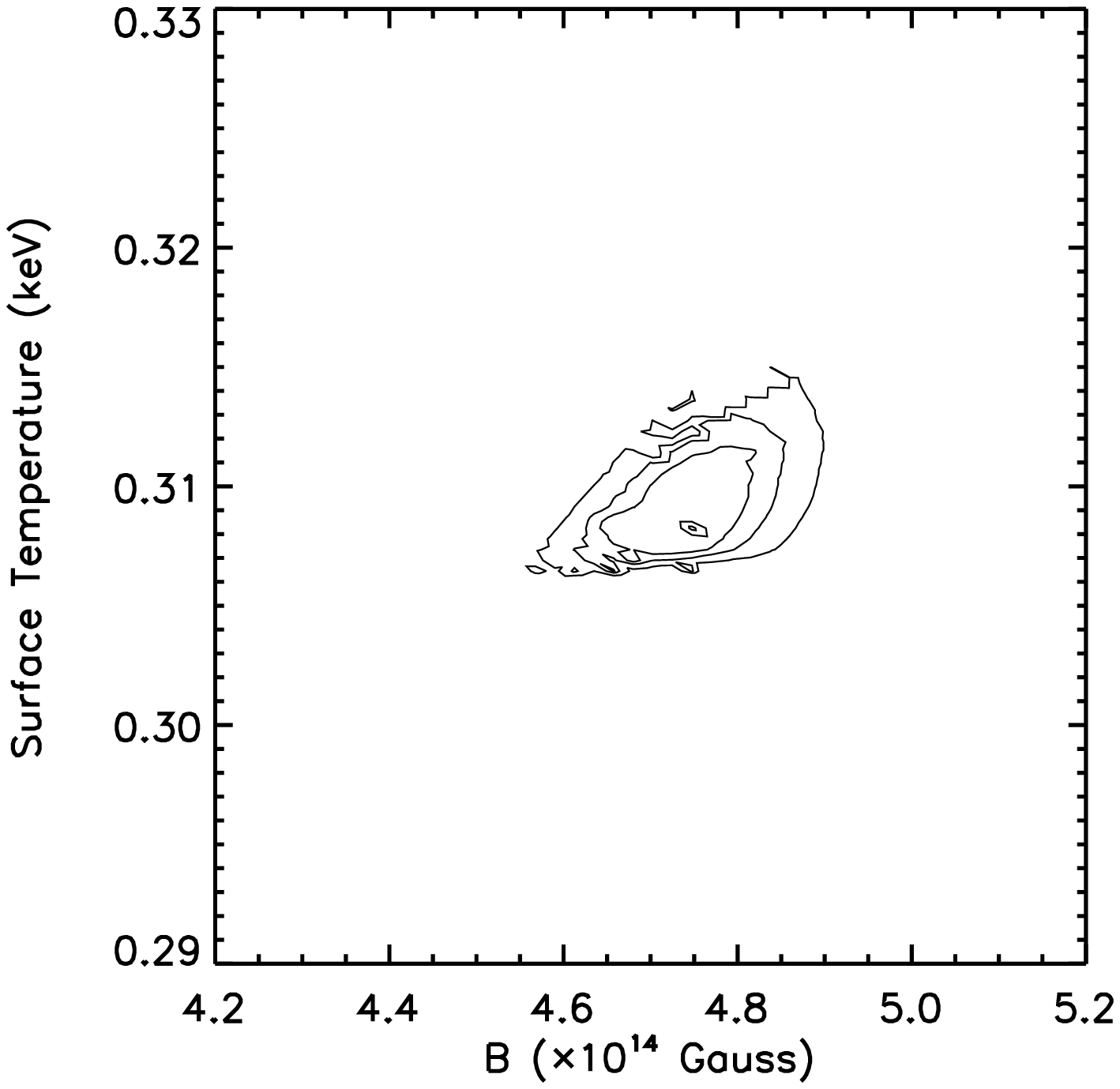}
   \includegraphics[scale=0.40]{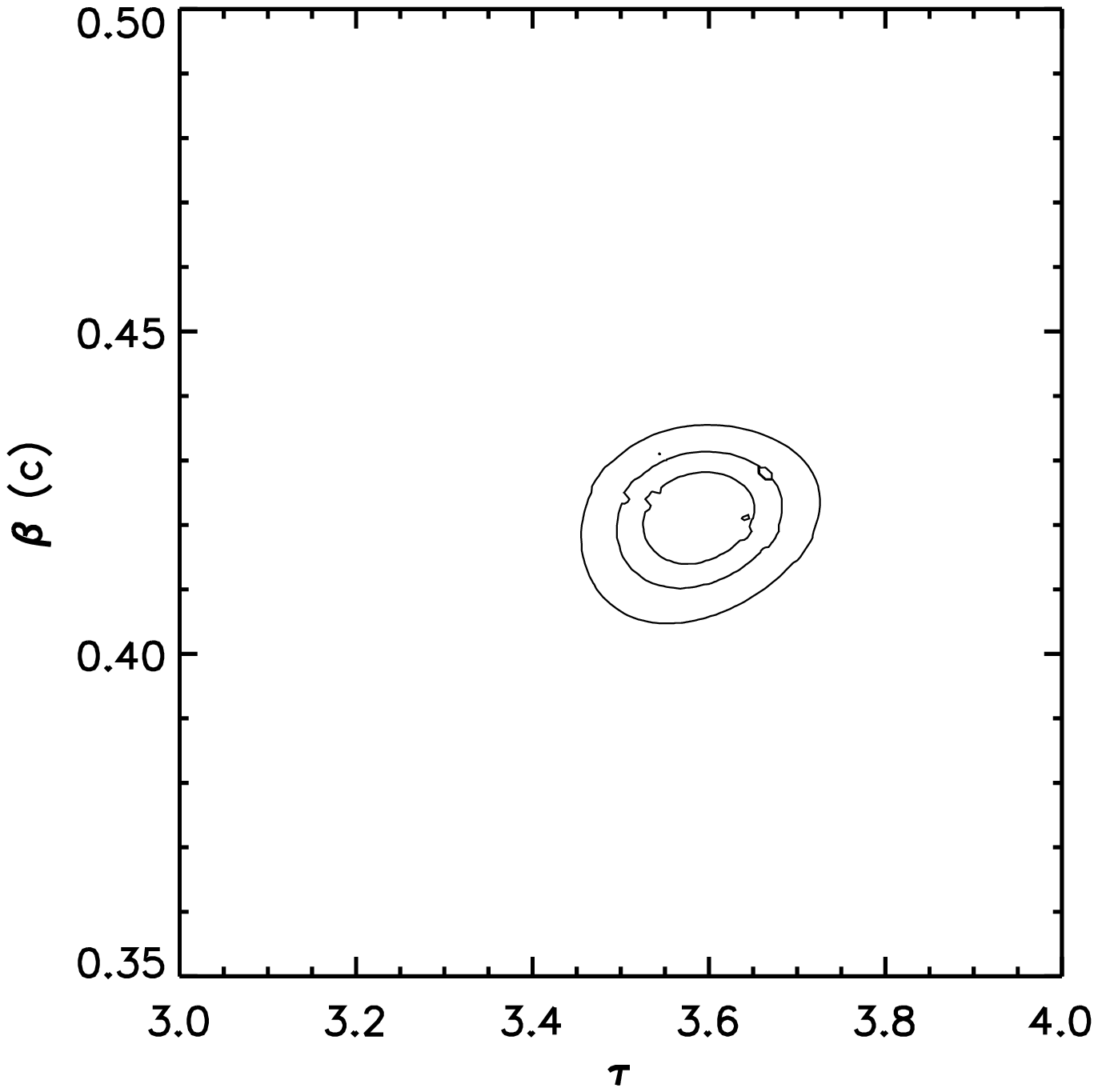}
   \includegraphics[scale=0.40]{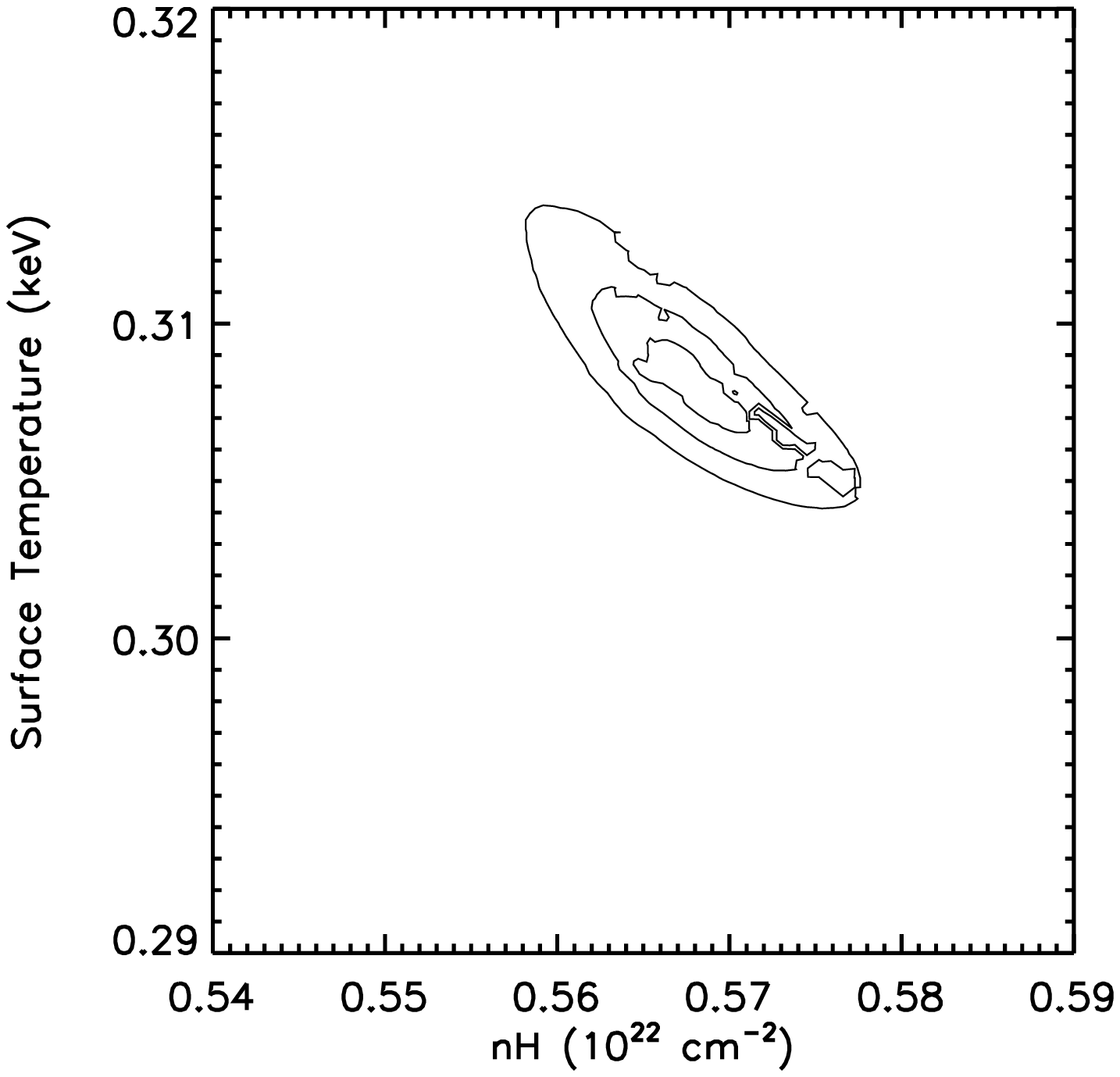}
   \includegraphics[scale=0.40]{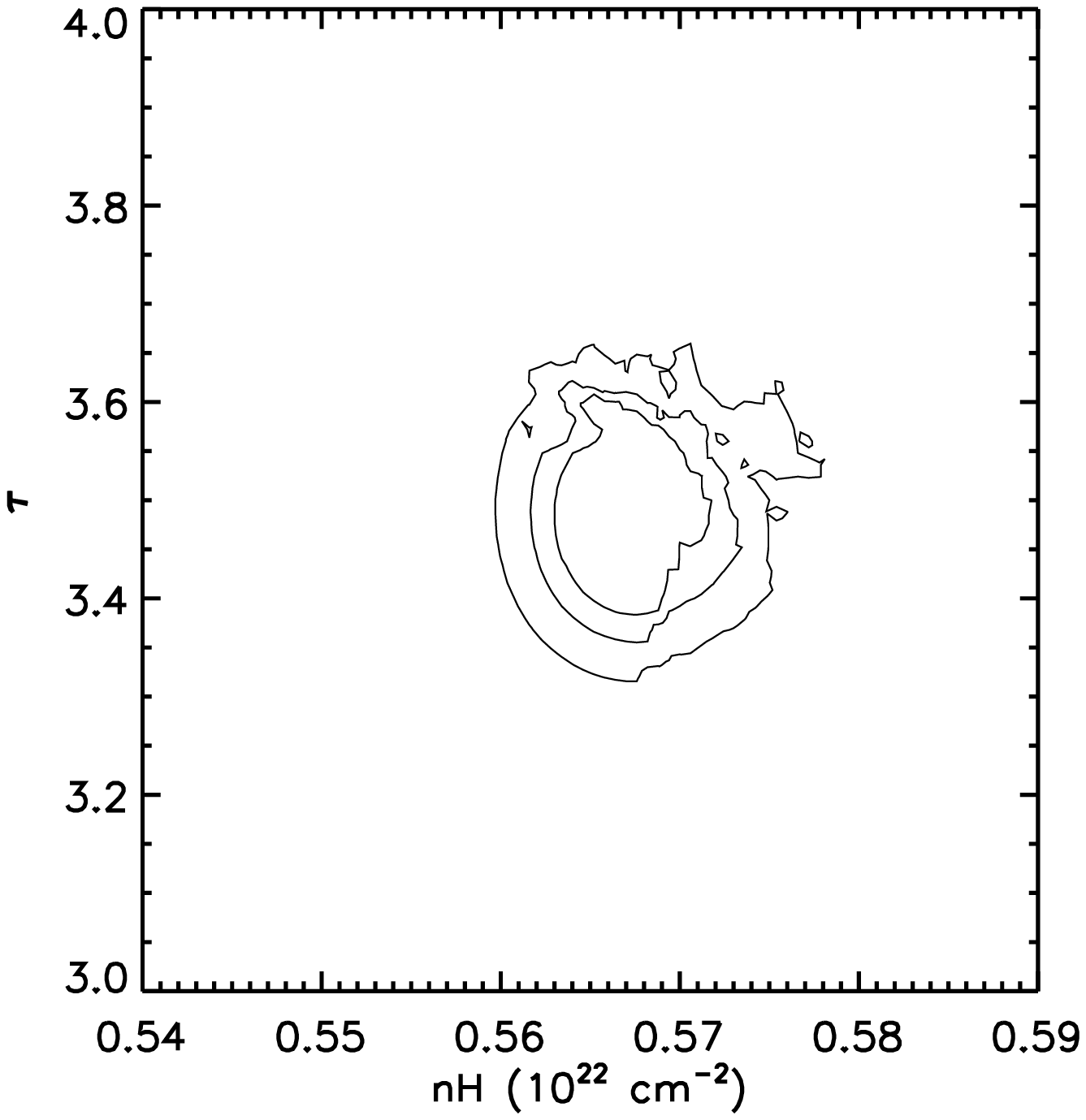}
   \caption{Confidence contour plots of different model parameters for
the fits shown in Figure~1. The three levels correspond to one-, two-,
and three-sigma confidence.}
\label{sim_conf}
\end{figure*}

\end{document}